# Predicting charge density distribution of materials using a local-environment-based graph convolutional network

Sheng Gong[1], Tian Xie[1], Taishan Zhu[1], Shuo Wang[2], Eric R. Fadel[1], Yawei Li[3], and Jeffrey C. Grossman[1*]

[1] *Department of Materials Science and Engineering, Massachusetts Institute of Technology, Cambridge, MA 02139, USA*

[2] *Department of Materials Science and Engineering, College of Engineering, Peking University, Beijing 100871, China*

[3] *Department of Chemical Engineering, Pennsylvania State University, University Park, PA 16802, USA*

## ABSTRACT

Electron charge density distribution of materials is one of the key quantities in computational materials science as theoretically it determines the ground state energy and practically it is used in many materials analyses. However, the scaling of density functional theory calculations with number of atoms limits the usage of charge-density-based calculations and analyses. Here we introduce a machine learning scheme with local-environment-based graphs and graph convolutional neural networks to predict charge density on grid-points from crystal structure. We show the accuracy of this scheme through a comparison of predicted charge densities as well as properties derived from the charge density, and the scaling is O($N$). More importantly, the transferability is shown to be high with respect to different compositions and structures, which results from the explicit encoding of geometry.

The electron charge density distribution is of enormous importance to the computational understanding and design of materials, as many fundamental properties relevant to a wide range of applications are directly related to the magnitude, shape, and variation of the charge density as well as its response to external stimuli. The charge density and its related properties, such as the electrostatic potential [1], electron localization function [2] and non-covalent interaction index [3], are directly used in analyses for many materials characteristics, including bonding [4], defects [5], stability [6], reactivity [7], and electron [8,9], ion [10,11] and thermal [12] transport, to name only a few. Recently, with the rapid development of machine learning (ML) applications in physics [13,14], chemistry [15,16] and materials science [17-19], charge densities are increasingly used as input features for predicting other materials properties in order to improve performance [20-22]. Currently the most common approach used to calculate charge density is density functional theory (DFT), which strikes a balance between accuracy and applicability. However, the relatively high computational cost and high memory demands of DFT [23] limits its use for large systems with more than several hundred atoms. Therefore, it is important to develop methods capable of accurately predicting charge density with less computational demand, to "by-pass the Kohn-Sham equations" [24], and ML is a promising tool for this goal due to the success of its application in predicting other DFT-computed properties [14,25-28].

In principle, an ideal ML algorithm should meet three requirements: high accuracy, high transferability and low computational cost [29]. Very recently, there have been attempts [24,30] to employ ML to predict the charge density of molecules by expanding the density as a sum of atom-basis functions. For the case of periodic systems, Schmidt *et al.*[31] employed basis functions, summing over the contributions from only neighboring atoms to achieve transferability between different cell sizes and lower memory demands, while Chandrasekaran *et al.*[23] encoded the position of each grid-point to neighboring atoms by a hierarchy of features with scalar, vector and tensor invariants to predict charge density. In both of these works the ML schemes were able to generate high quality charge densities with O($N$) scaling, although compositional and structural transferability remains a challenge, as these methods account for variations in one structure at a time (i.e., strained lattices or different molecular dynamics snapshots). While these approaches have shown early promise in the development of ML algorithms for charge density prediction, there remains a need for ML-based methods that can efficiently and accurately be applied to structures with different elemental compositions and structural features.

Here, we develop a ML-based approach that can predict charge density for different structures with varying compositions, structural features and defects for a given class of materials in a single training, which is necessary for application to systems such as amorphous hydrocarbons or glasses where local structures are highly complex. In previous works, a three-step process was followed: 1) record the distance between each grid point and all neighboring atoms, 2) add all distances together to form a feature vector, and 3) compute charge density by regression on the final feature vector. For multi-elemental systems, the first two steps are repeated for each element type and the feature vectors are concatenated together. The success of this approach shows that the charge density distribution in a single structure can be sufficiently learned by the sum of contributions from neighboring atoms.

In order to build upon this approach with the aim of increasing transferability between different structures, in addition to recording the distance between grid-points and atoms, we propose to both explicitly encode the geometry of the cluster formed by neighboring atoms, and account for all elements simultaneously as opposed separately. Encoding the geometry, on the one hand, avoids the problem of different local environments leading to a similar sum of atom contributions (FIG. S1(a)), on the other hand, enables the model to learn from the geometry of existing structural features and speculate new ones (FIG. S1(b)). A similar idea is discussed in Schmidt *et al.* [31] by considering contributions of atom-pairs. Greater structural transferability should also lead to improved accuracy in the prediction of charge density for defect structures, as new structural features can form during the formation of defects. To accommodate different elements, the dimension of the final feature vector should be independent of composition, otherwise the regression process (matrix-vector multiplication) cannot be done for feature vectors with different dimensions.

A graph representation, which encodes both nodes and bonds, has a number of advantages that meet the requirements listed above. Graph representations have been used recently to encode information on both the level of atom and geometry with high accuracy and transferability across composition, structure and property space [14,17], and the feature vectors can be of the same dimension for different compositions if properly designed. In this work, we encode environments of grid-points as graphs and employ the crystal graph convolution neural network [14] (CGCNN) to find a relationship between local environment and charge density with O($N$) scaling. We train and test our scheme on two classes of crystalline materials, polymers and zeolites. For each case training data is used from some structures and the model is applied to others in order to test transferability, and the accuracy of the predicted charge density is evaluated through statistics, visualization and accuracy of its derivative and related properties (i.e., dipole moment).

We encode three dimensional space in the unit cell using CGCNN by placing an imaginary atom at each grid-point in the unit cell (FIG. 1). The local environment is computed for a given grid-point by identifying atoms within a cut-off radius ($R_{cut}$) from the imaginary atom, as shown in FIG. 1(b). Next as shown in FIG. 1(c), atoms outside $R_{cut}$ are removed, and the remaining structure is placed in a larger cell to avoid interactions between periodic images. Here $R_{cut}$ is set to be 4 Å, which is larger than typical bond lengths for the materials considered in this work [32], and the lattice parameters of the larger cell are set to be no less than $3\times R_{cut}$. Finally, the remaining structure together with the imaginary atom are converted into a graph representation as shown in FIG. 1(d) by connecting neighbors. The CGCNN is then trained on the local-environment-based graphs with the charge density on the grid-points from DFT calculations as the target property (with units of e/Å$^3$). Details of the DFT calculations and representation of the imaginary atom are given in Supplementary Materials.

The complete framework of CGCNN is presented in Ref. [14] so here only a brief description is provided. The neural network structure is summarized in FIG. 1(d). Once given a graph, the convolutional layers iteratively update the atom feature vector $\boldsymbol{v}_\mathbf{i}$ based on surrounding atoms and bonds with a convolution function:

$$\boldsymbol{v}_\mathbf{i}^{\mathbf{(t+1)}} = \mathrm{Conv}(\boldsymbol{v}_\mathbf{i}^{\mathbf{(t)}}, \boldsymbol{v}_\mathbf{j}^{\mathbf{(t)}}, \boldsymbol{\mu}_\mathbf{ij}) \qquad (1),$$

where $\boldsymbol{v}_{\mathbf{i(j)}}^{\mathbf{(t)}}$ is the atom feature vector of the i(j) th atom after t convolutions, $\boldsymbol{\mu}_\mathbf{ij}$ represents the bond vector between the i th and j th atoms and Conv stands for the convolution function. Here the convolution function designed in Ref. [14] is used, which was shown to be accurate for encoding interaction strengths and produces feature vectors with constant dimension for different compositions. A pooling function is then used to create an overall feature vector to satisfy permutational and size invariance as:

$$\boldsymbol{v} = \mathrm{Pool}(\boldsymbol{v}_\mathbf{0}^{\mathbf{(0)}},\ldots, \boldsymbol{v}_\mathbf{0}^{\mathbf{(T)}},\ldots, \boldsymbol{v}_\mathbf{N}^{\mathbf{(T)}}) \qquad (2).$$

Here, the mean of atom vectors is taken as the feature after pooling for simplicity, while other pooling functions can also be used.

In addition to convolution and pooling, two hidden layers are used to capture the complex relationship between structure and property, and finally an output layer is used to give the target property. This process meets both of the requirements as mentioned above, since after convolution the atom feature vector for the imaginary atom encodes the distances between one grid-point and neighboring lattice atoms, while that for lattice atoms encodes their position with respect to not only other lattice atoms but also the imaginary atom. The pooling process incorporates all the information together, making the final feature vector informative and of the same dimension for materials with different compositions.

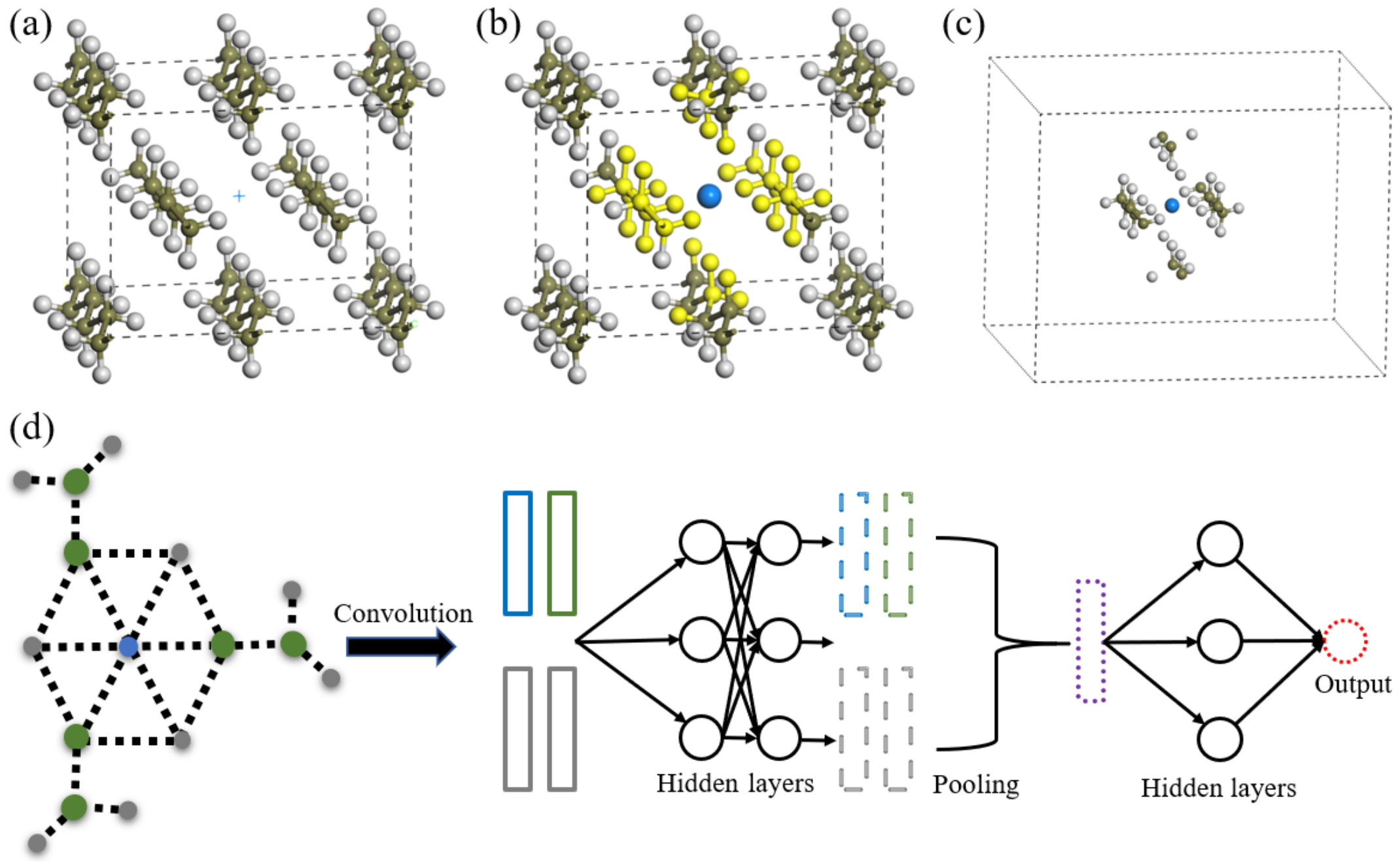

**FIG. 1.** Procedure of converting the local environment into a graph, using ethylene as an example. (a) Crystal structure of crystalline ethylene. The blue plus symbol in the center denotes a grid point we are interested in. (b) Crystalline ethylene with the imaginary atom. Highlighted atoms are those within the cut-off radius. (c) Local environment around the imaginary atom. (d) Sketch of local-environment-based graph and CGCNN architecture. Color coding: green: carbon; grey: hydrogen; blue: imaginary atom; yellow: highlighted atoms within the cut-off radius.

In the case of crystalline polymers, we extract 30000 graphs from 37 different structures as training data, while in the case of zeolites 8000 graphs are generated from 5 different structures for training. The list of structures from which training data are obtained is provided in Table S2. Further details related to dataset construction and grid spacing are provided in Supplementary Materials.

In FIG. 2(a), we show how the prediction performance changes as a function of the training size for the polymer and zeolite materials considered. The straight-line-like trends in FIG. 2(a) indicate that better performance is possible with larger training sets. In addition, the steeper slope for the case of zeolites indicates their reduced chemical complexity compared to the polymers, which is discussed further below. As for the computational cost, although direct comparison between computation time of DFT and ML is difficult as they are based on different computing architectures, in FIG. 2(b) the relation between computational time and number of atoms in the system is plotted for prediction of the charge density of crystalline p-xylylene using our ML model and DFT calculations, from which one can see the linear scaling of the ML approach.

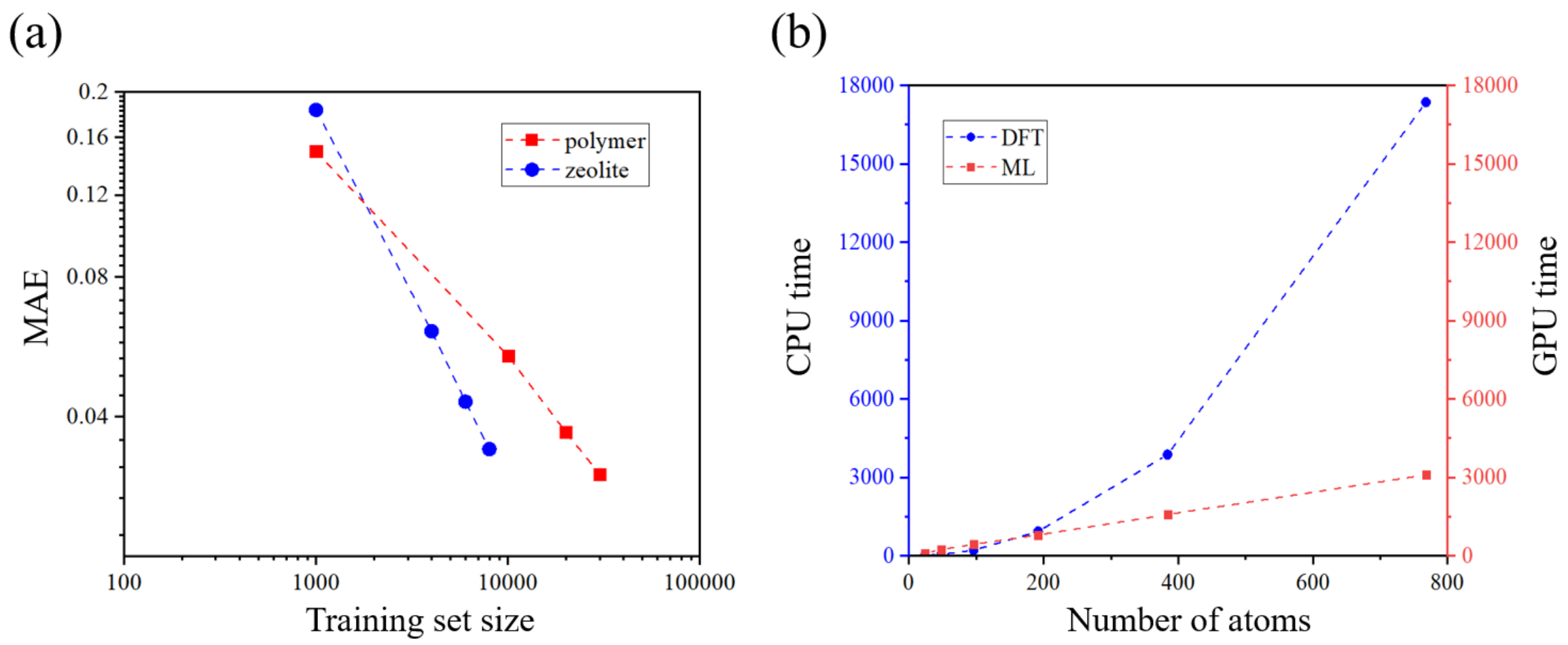

**FIG. 2.** (a) Mean average error (MAE, in e/Å$^3$) of the ML predicted charge density of the test sets (grid-points) from the training structures versus training set size for polymer and zeolite materials. (b) CPU time (in seconds) for DFT calculations and GPU time (in seconds) for ML prediction versus number of atoms in the cell for crystalline p-xylylene. DFT calculations are performed by 24 Intel(R) Xeon(R) CPUs with RAM of 128GB, while ML calculations are carried out on a single NVIDIA GeForce GTX 1070 GPU with RAM of 2 GB.

In order to test the degree of transferability towards different structures, we apply our model to predict the charge density of 17 crystalline polymers and 9 zeolites not included in the training sets (see Table 1). In both cases, the nomex polymer and NPO zeolite, also have versions with explicitly created defect structures (denoted as nomex_defect and NPO_defect) in order to represent additional chemical complexity. These materials are not subsets of the training sets in terms of structure or size. Structural features are represented by coordinations of skeleton atoms (C/O in the case of polymer/zeolite). For example, C2H2 means there are 2 C atoms and 2 H atoms coordinated with the central atom. For polymers, in FIG. 3(a) the frequency of different coordinations for carbon atoms is shown for both the training and test sets, from which one can see that nearly 20 different coordinations appear, showing considerable bonding complexity. More importantly, there are three coordinations in the test set that are not included in the training set (H4, C1H1 and C4). For zeolites, the training set is simpler than the polymer set in terms of structure, as only two coordinations exist, and in the test set only the structure with a defect, NPO_defect, has the coordination of Si1, while all other structures have coordination Si2. Further details regarding the chemical complexity of the datasets based on composition and size are provided in the Supplementary Materials.

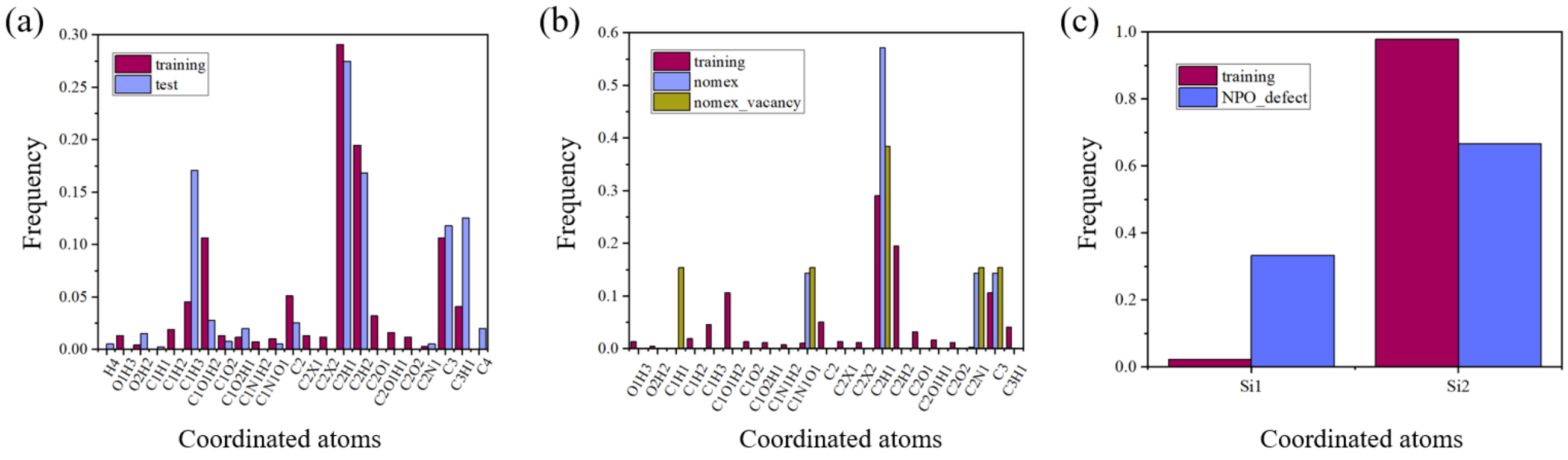

**FIG. 3.** (a) and (b) Appearance frequency of coordinated atoms of carbon atoms in the training set for the case of crystalline polymers versus the test set as a whole and nomex and nomex_defect, respectively. Here 'X' denotes rare elements in our case (Cl, F, S, Si, Hg), and as an example, C2H2 means there are 2 C atoms and 2 H atoms coordinated with the central atom.. (c) Appearance frequency of oxygen coordinated atoms in the training set for the case of zeolites versus the structure of NPO_defect.

Here, we choose two metrics, root mean square errors (RMSE) and coefficients of determination ($R^2$), to quantify errors in the ML predicted charge density. These metrics, which are also used in Schmidt *et al.*[31], provide insights on both the magnitude of absolute errors (by RMSE) and relative performance of the predictions (by $R^2$). As shown in Table 1, the RMSE of the predicted charge densities are less than 0.1 e/Å$^3$, which are comparable to the errors in Ref. [31], and the level of accuracy was demonstrated to be sufficient for most applications relying on the accuracy of the density representation [33]. The RMSEs of test structures are also close to that of the training sets (0.067 e/Å$^3$ and 0.064 e/Å$^3$ for crystalline polymers and zeolites, respectively), indicating little overfitting. More importantly, the $R^2$ are larger than 0.95 for all test structures, suggesting a high prediction performance. The results for the case of zeolites show that for such a simple materials class, accurate prediction of the charge density can be achieved with a relatively small training set (less than 10,000 training data in this case). In addition to these general trends, we highlight the cases of i-4m1p, isobutylene, and the

nomex_defect, which possess different coordination environments. Although larger errors are observed in these cases, they are not far from other structures with RMSE < 0.1 e/Å$^3$ and $R^2 > 0.95$, suggesting good transferability to unseen structural features.

**TABLE 1.** Root mean square errors (RMSE) and coefficients of determination ($R^2$) of the ML predicted charge density (ρ, in e/Å$^3$) and Laplacian of charge density ($\nabla^2\rho$, in e/Å$^5$). For each structure, the error metrics are computed over all grid-points in the unit cell. The last nine structures with 3-letter abbreviations are zeolites, and others are crystalline polymers.

| name | formula (inside the cell) | RMSE (ρ) | $R^2$ (ρ) | RMSE ($\nabla^2\rho$) | $R^2$ ($\nabla^2\rho$) |
|---|---|---|---|---|---|
| 1,3-dioxolane-II | $C_{24}H_{48}O_{16}$ | 0.0628 | 0.9933 | 0.4190 | 0.9934 |
| acetaldehyde | $C_{32}H_{64}O_{16}$ | 0.0818 | 0.9848 | 0.5007 | 0.9850 |
| cis-1,4-butadiene | $C_{16}H_8$ | 0.0902 | 0.9805 | 0.3502 | 0.9822 |
| glycolide | $C_8H_8O_8$ | 0.0681 | 0.9943 | 0.4502 | 0.9941 |
| gutta-percha-alpha | $C_{20}H_{32}$ | 0.0369 | 0.9953 | 0.1998 | 0.9939 |
| i-4m1p | $C_{168}H_{336}$ | 0.0666 | 0.9729 | 0.4521 | 0.9656 |
| i-alpha-vnaph | $C_{192}H_{160}$ | 0.0661 | 0.9816 | 0.4311 | 0.9778 |
| i-ortho-mths | $C_{144}H_{160}$ | 0.0593 | 0.9831 | 0.3678 | 0.9798 |
| i-propylene-alpha | $C_{36}H_{72}$ | 0.0491 | 0.9881 | 0.2992 | 0.9852 |
| isobutylene | $C_{64}H_{128}$ | 0.0910 | 0.9569 | 0.6114 | 0.9541 |
| nomex | $C_{14}H_{10}O_2N_2$ | 0.0626 | 0.9926 | 0.3333 | 0.9899 |
| nomex_defect | $C_{13}H_9O_2N_2$ | 0.0665 | 0.9913 | 0.3590 | 0.9882 |
| oxymethylene | $C_4H_8O_4$ | 0.0786 | 0.9926 | 0.4765 | 0.9906 |
| p-xylylene | $C_{16}H_8$ | 0.0580 | 0.9890 | 0.2735 | 0.9916 |
| s-propylene-1 | $C_{24}H_{12}$ | 0.0523 | 0.9835 | 0.3359 | 0.9814 |
| tetramtht | $C_{12}H_{12}O_4$ | 0.0502 | 0.9960 | 0.3538 | 0.9954 |
| trans-decenamer | $C_{10}H_{18}$ | 0.0309 | 0.9970 | 0.3590 | 0.9882 |
| NPO | $Si_6O_{12}$ | 0.0977 | 0.9893 | 0.5602 | 0.9885 |
| NPO_defect | $Si_5O_{12}$ | 0.0998 | 0.9845 | 0.5989 | 0.9821 |
| JBW | $Si_6O_{12}$ | 0.0847 | 0.9914 | 0.5702 | 0.9887 |
| CAN | $Si_{12}O_{24}$ | 0.0831 | 0.9906 | 0.6014 | 0.9893 |
| AFY | $Si_{16}O_{32}$ | 0.0778 | 0.9894 | 0.5418 | 0.9879 |
| JSN | $Si_{16}O_{32}$ | 0.0785 | 0.9911 | 0.5221 | 0.9901 |
| MTN | $Si_{136}O_{272}$ | 0.0821 | 0.9903 | 0.2809 | 0.9886 |
| TUN | $Si_{192}O_{384}$ | 0.0754 | 0.9920 | 0.1986 | 0.9922 |
| UOV | $Si_{176}O_{352}$ | 0.0912 | 0.9881 | 0.2039 | 0.9876 |

Next, the Laplacian of the charge density is computed in order to test the ML model's ability to capture variation in charge. The Laplacian of the charge density is of great importance to functional construction [34] and materials analysis [35], and from Table 1 we can see that the Laplacian is also well predicted with $R^2 > 0.95$.

In order to visualize the performance and transferability of our model, we compare the ML computed charge densities and difference between charge densities from ML and DFT of pristine nomex, nomex with a carbon-hydrogen vacancy, pristine NPO and NPO with a Si vacancy in FIG. 4. In all the cases, the building blocks of structures (e.g., the C six-ring and Si-O six-ring) are well presented. For defect structures, although there are more significant differences between ML and DFT, the magnitude of the difference is still low compared with the charge density itself, suggesting high transferability towards defect structures.

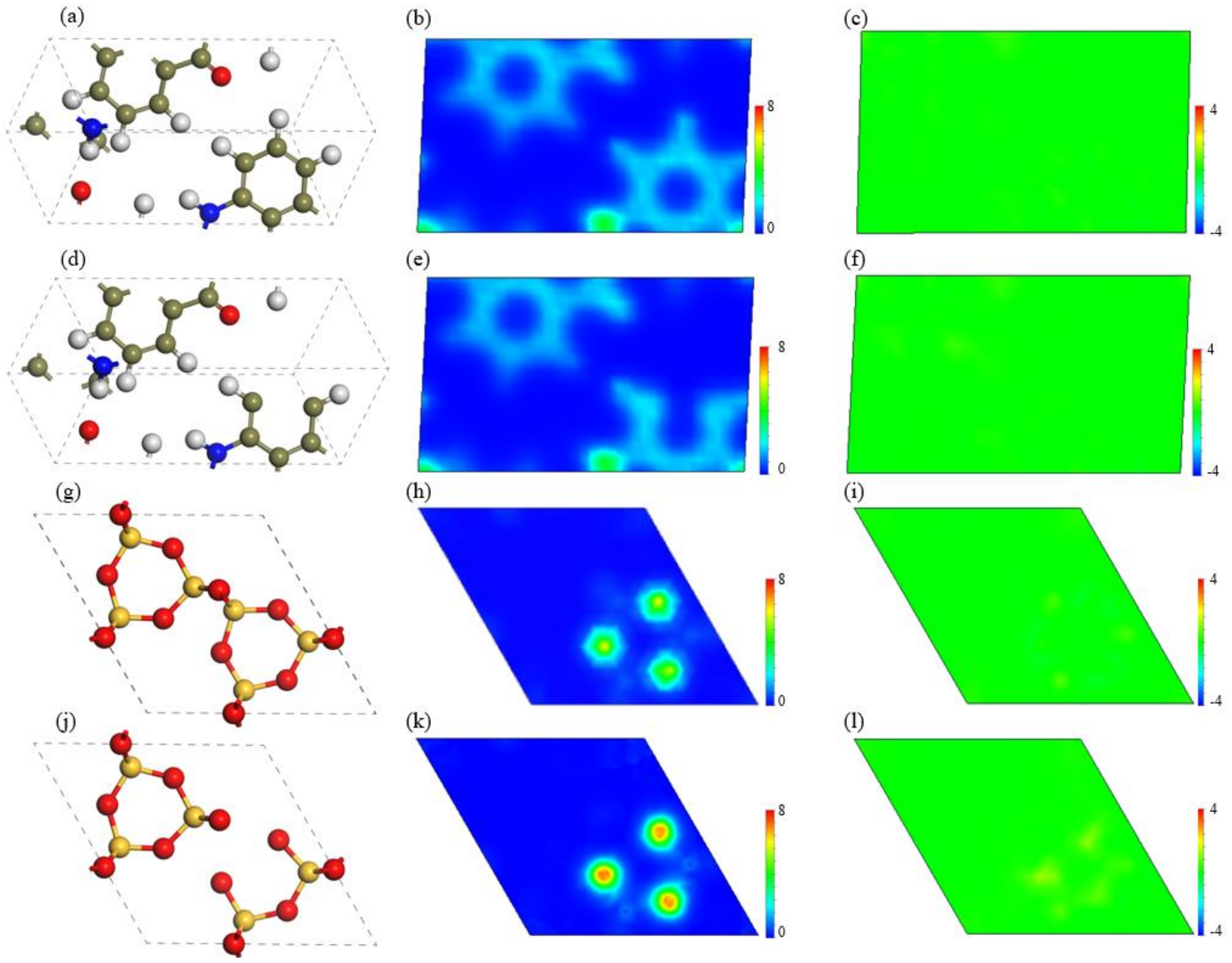

**FIG. 4.** Visualization of electron charge density (ρ, in e/Å$^3$). (a), (b), (c) and (d), (e), (f) crystal structure, ML predicted ρ, and difference between ML predicted ρ and DFT calculated ρ on the C six-ring plane of pristine nomex and nomex with a carbon and a hydrogen vacancy, respectively. (g), (h), (i) and (j), (k), (l) crystal structure, ML predicted ρ, and difference between ML predicted ρ and DFT calculated

ρ on the Si-O six-ring plane of pristine NPO and NPO with a Si vacancy, respectively. Atom color coding: green: carbon; grey: hydrogen; red: oxygen; blue: nitrogen; yellow: silicon.

We further compare the value of ML predicted ρ versus DFT calculated ρ as shown in FIG. 5. The ML model successfully captures the charge densities in most regions for the four structures with high $R^2$. As shown in FIG. 5(b) and (d), our ML model is able to accurately capture the charge density of a vacancy even though no defect structures were present in the training sets. Meanwhile, we can see that most of the deviation in the ML approach compared with DFT is from regions with ultrahigh charge density (near atom cores as shown in FIG. 4), which offers insight into directions for further improvement as discussed below.

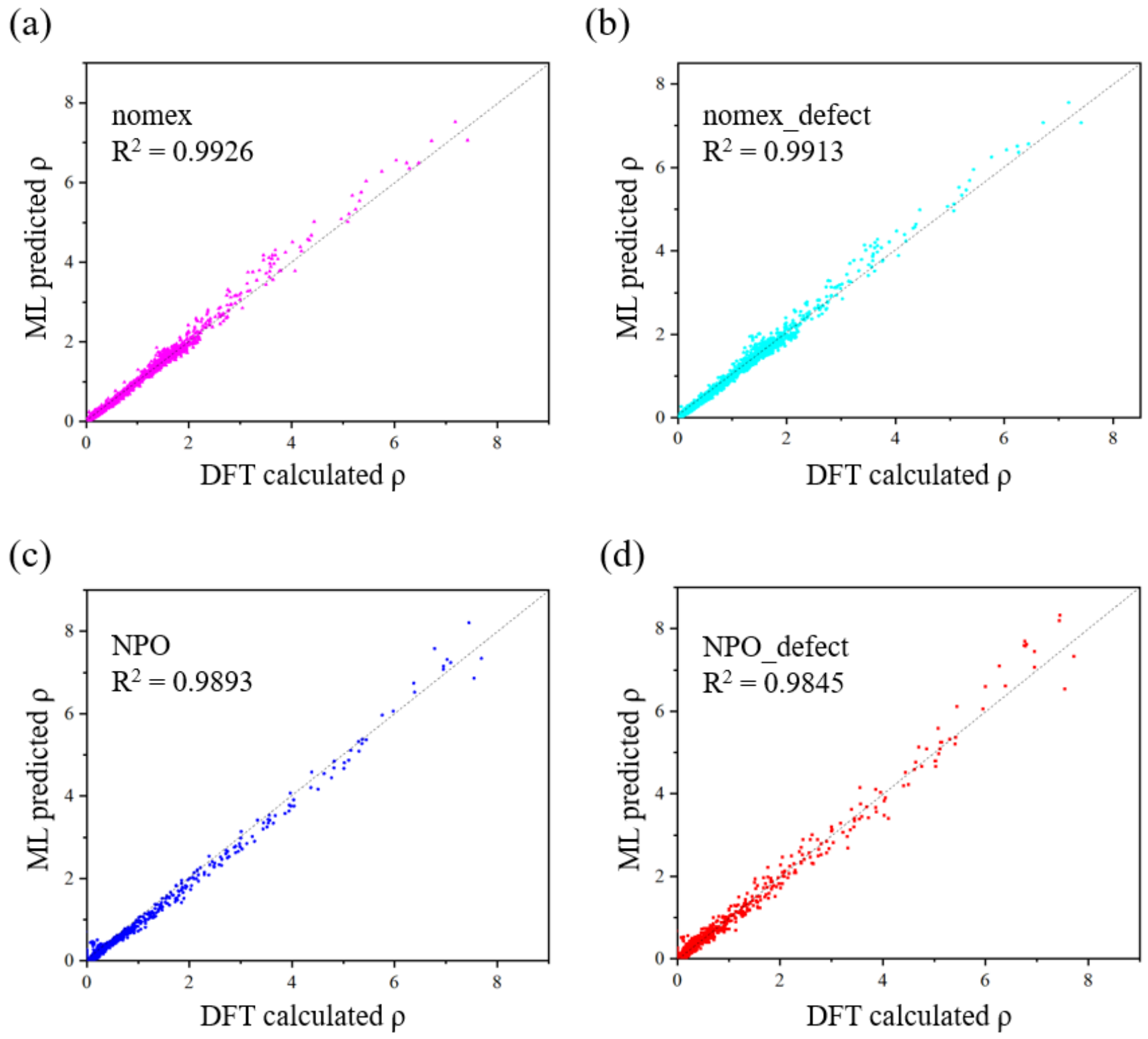

**FIG. 5.** (a), (b), (c) and (d) ML predicted charge density (ρ, in e/$Å^3$) versus DFT calculated ρ for pristine nomex, nomex_defect, pristine NPO and NPO_defect, respectively.

Last, we evaluate the accuracy of our model for predicting the dipole moment, which is a materials property that can be derived from the charge density. Here we use the predicted charge density of half of the test sets (13 structures) to derive dipole moments for the unit cells, generated from the crystal structure and charge distribution as:

$$\mathbf{v_e} = \int_{cell} \mathbf{r}\bullet\rho(\mathbf{r})d\mathbf{r};\ \mathbf{v_i} = \int_{cell} \mathbf{r}\bullet Z(\mathbf{r})d\mathbf{r};\ \mu = |\mathbf{v_e}+\mathbf{v_i}|/V_{cell} \qquad (3),$$

where $\mathbf{r}$ denotes position vector, $V_{cell}$ is the volume of cell, $\rho(\mathbf{r})$ and $Z(\mathbf{r})$ are charges from electron and ion (opposite sign) on $\mathbf{r}$, $\mathbf{v_e}$ and $\mathbf{v_i}$ represent electron and ion dipole vectors, and $\mu$ is the dipole moment per volume in the unit cell, respectively. The results are shown in Table 2, and we can see that the differences between the two electron dipole vectors (from ML and DFT) are very small in all the cases with a high $R^2$ of 0.99. As for the total dipole moment, although comparative deviations increase after cancellation of contribution from positive and negative charge, our model can still achieve a $R^2$ of 0.89, which is close to that of machine learning schemes designed specifically for dipole moments (0.93 in Pereira *et al.*[36] and 0.91 in Bereau *et al.*[37]).

**TABLE 2.** Electron dipole vectors ($\mathbf{v_e}$, in e•Å) and total dipole moment ($\mu$, in Debye/Å$^3$) from ML predicted $\rho$ and DFT calculated $\rho$ in the unit cells of half the test structures, respectively.

| name | $\mathbf{v_e}$ (ML) | $\mathbf{v_e}$ (DFT) | $\mu$ (ML) | $\mu$ (DFT) |
|---|---|---|---|---|
| nomex | (20.9, 20.0, 431.4) | (19.1, 19.2, 422.3) | 0.652 | 0.781 |
| nomex_defect | (31.9, 27.5, 440.3) | (29.9, 26.9, 432.0) | 0.647 | 0.782 |
| s-propylene-1 | (858.7, 323.8, 438.0) | (898.9, 339.8, 458.5) | 0.141 | 0.260 |
| glycolide | (219.0, 273.9, 295.4) | (216.1, 271.2, 291.8) | 0.366 | 0.463 |
| p-xylylene | (38.6, 363.9, 161.0) | (38.0, 362.6, 160.4) | 1.095 | 1.076 |
| tetramtht | (80.0, 78.8, 381.5) | (83.2, 81.8, 396.5) | 0.572 | 0.383 |
| trans-decenamer | (42.6, 234.5, 306.2) | (44.0, 243.2, 317.5) | 0.649 | 0.409 |
| NPO | (185.4, 321.2, 220.3) | (196.7, 340.6, 232.8) | 1.025 | 1.248 |
| NPO_defect | (202.4, 333.0, 229.0) | (184.4, 325.6, 226.7) | 1.390 | 1.322 |
| JBW | (240.6, 349.7, 362.6) | (232.7, 341.2, 350.4) | 1.799 | 1.547 |
| CAN | (572.9, 992.4, 466.0) | (573.9, 994.1, 465.2) | 1.294 | 1.302 |
| AFY | (787.4, 1363.8, 1094.4) | (777.0, 1345.8, 1081.6) | 1.247 | 1.143 |
| JSN | (857.1, 827.1, 1855.7) | (866.6, 831.6, 1869.2) | 1.170 | 1.260 |

The difficulty of transferability between different structures arises from both training and prediction: in training, the model has to distinguish between environments that seems to be ‘similar’ but have very different values of charge, and in prediction, the model has to find similarities between new and existing features. Here, the geometry of neighboring atoms contained in our graph representation simultaneously provides the information for the two tasks, which leads to the improved transferability of our model. On the one hand, encoding the geometry makes the local environments more distinguishable as shown in FIG. S1(a); on the other hand, learning the geometry enables the model to speculate new structural features from existing ones. For example, for the coordination of C4, although it is not in the training set, as shown in FIG. S1(b) the model can learn from the tetrahedral geometries of C1H3, C2H2 and C3H1 that the central carbon atoms are $sp^3$ hybridized, which facilitates prediction of charge density around the central carbon atom. Encoding the geometry also helps to predict the shape of charge density around the defects from the shape of structural features, as illustrated in FIG. S1(c).

Future efforts will be applied to further improve the scheme presented in three aspects. First, as mentioned we will design architectures to efficiently generate more materials properties based on charge density, especially the total energy of the unit cell, for which both traditional methods (e.g. Kohn-Sham equations [38] or embedded-atom method [39]) and machine learning approaches [21,22] are options under consideration. Second, as discussed above regions near nuclei possess the highest deviations, and to improve the sensitivity of our model for small distances between imaginary and real atoms, transformations to weight small distances during the learning can be designed. Third, although here $R_{cut}$ = 4 Å works well for the example cases studied, for systems where long range interactions are important the efficiency of our model will drop fast. For such cases we suggest that a series of tests should be carried out to determine the optimal $R_{cut}$, and in the future physical insights will be used to determine the relationship between the optimal $R_{cut}$ and interaction mechanism for different materials systems.

In summary, we have developed a machine learning model to predict electron charge density distribution of materials based on graph convolutional neural networks with O($N$) scaling. In the case studies of crystalline polymers and zeolites, local-environment-based graphs are extracted from some structures and features learned, and applied to structures different from the training sets. The accuracy and usability of our model has been evaluated by statistical errors, visualization and quality of related quantity and property of charge density. The most important benefit of our model is high transferability between different structures, which can be attributed to the ability of the graph representation to explicitly encode the geometry of neighboring atoms for each grid-point.

This work was supported by Toyota Research Institute. Computational support was provided by the DOE Office of Science User Facility supported by the Office of Science of the U.S. Department of Energy under Contract No. DE-AC02-05CH11231, and the Extreme Science and Engineering Discovery Environment, supported by National Science Foundation grant number ACI-1053575.

# Predicting charge density distribution of materials using a local-environment-based graph convolutional network

Sheng Gong[1], Tian Xie[1], Taishan Zhu[1], Shuo Wang[2], Eric R. Fadel[1], Yawei Li[3], and Jeffrey C. Grossman[1*]

[1] *Department of Materials Science and Engineering, Massachusetts Institute of Technology, Cambridge, MA 02139, USA*

[2] *Department of Materials Science and Engineering, College of Engineering, Peking University, Beijing 100871, China*

[3] *Department of Chemical Engineering, Pennsylvania State University, University Park, PA 16802, USA*

## 1. Discussions about encoding geometry of neighboring atoms

In order to illustrate the impact of encoding the geometry of neighboring atoms for distinguishing local environments, we sketch two local environments in FIG. S1(a). If the environments of grid points are described by considering distances to each atom separately and then summing atom contributions as in the current models, the two environments would appear to be very similar. However, they are actually quite different, and the difference can be explicitly encoded by the distance between the two atoms, highlighting the importance of encoding atomic orders.

For speculating new structural features from existing ones, we plot the geometries of central carbon atoms with coordinated C1H3, C2H2, C3H1 and C4 atoms in FIG. S1(b). When predicting charge density around C4, our model can learn from the geometries of C1H3, C2H2, C3H1 in the training set that the tetrahedral shape of C4 corresponds to a *sp*$^3$-hybridized central carbon atom, which gives key information for charge distribution around the central carbon atom. As shown in FIG. 3(b), FIG. 4(d) and FIG. S1(c), in nomex_defect there is a structural feature (C1H1) that is formed during defect formation and doesn't exist in the training set with all pristine structures. However, as shown in FIG. S1(c), the shape of C1H1 (C-C-H, an obtuse angle) is very similar to that of C-O-H in the training set. Therefore, the charge distributions around the two structural features should be both in a shape of obtuse angle. With the information of geometries, our model can capture such similarity and predict the obtuse-angle-like charge density around C1H1, and the ratio of charge density between C-C and C-H atoms can be learnt from the 20+ structural features listed in the main text.

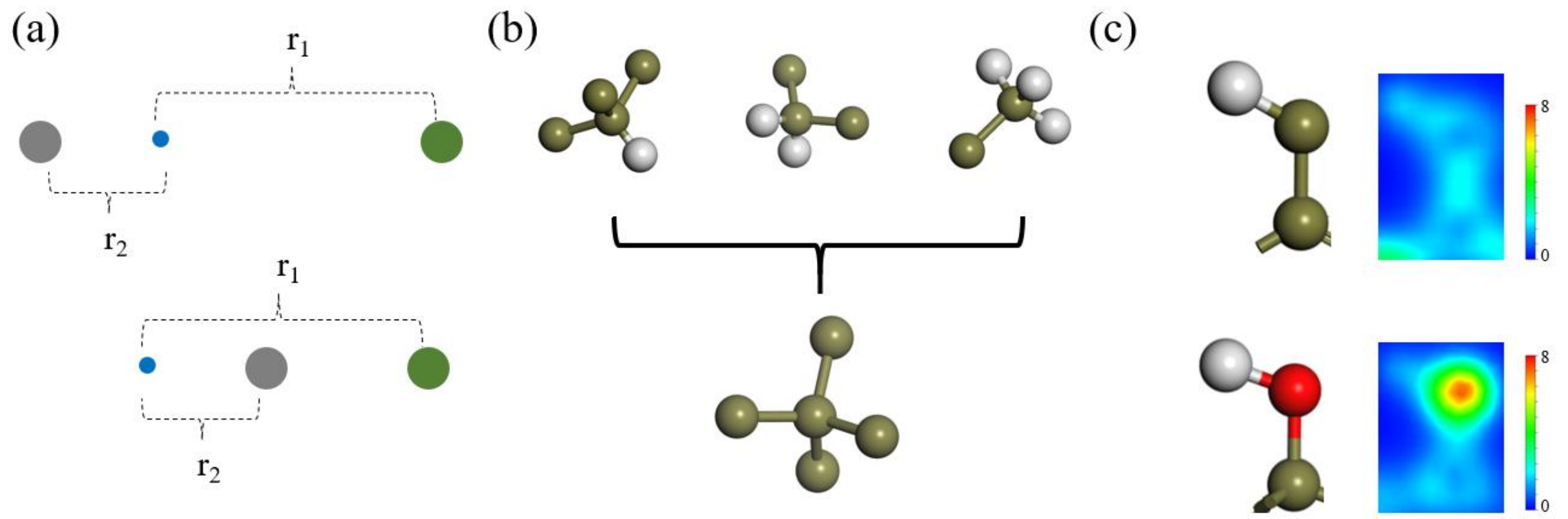

**FIG. S1.** (a) Sketch of two different local environments with similar sum of atom contributions. (b) Geometries of central carbon atoms with coordinated C1H3, C2H2, C3H1 and C4 atoms. (c) Shape of C-C-H and C-O-H and their charge density distributions (ρ, in e/Å$^3$). Atom color coding: green: carbon; grey: hydrogen; red: oxygen; blue: nitrogen.

## 2. Details of DFT calculations

DFT calculations to obtain charge density distributions are implemented in the Vienna *Ab initio* Simulation Package (VASP) [1]. The projector augmented wave (PAW) [2] scheme is used to treat the interactions between ion cores and valence electrons. The exchange-correlation is approximated by Perdew-Burke-Ernzerh functional (PBE) [3]. For the calculation of time scaling, the first Brillouin zone is sampled by a 2 × 2 × 2 k-point grid, while that for other calculations is of ~0.5 $Å^{-1}$. In order to account for van der Waals forces, the DFT-D2 [4] dispersion-correlation is used.

## 3. Discussion about how to represent the imaginary atom

In principle, any representation of the imaginary atom that is different from those for elements in our system is acceptable. Since CGCNN constructs a representation for atoms based on elemental properties, here for simplicity we use the representation of the He atom in CGCNN to represent the imaginary atom, as He doesn't exist in our cases nor most periodic systems, and Table S1 in Supplementary Information shows that different representations of the imaginary atom would lead to very similar performance. Nevertheless, when necessary one can always construct other representations different from all existing elements such as additional dimensions to tag the imaginary atom.

**TABLE S1.** Mean average error (MAE, in e/$Å^3$) of the training set in the zeolite case versus the choice of representation of imaginary atom.

| Choice of imaginary atom | He | Li | Ne | Cs | Xe |
|---|---|---|---|---|---|
| MAE | 0.030 | 0.041 | 0.037 | 0.035 | 0.036 |

## 4. Datasets construction and grid spacing

For the case of crystalline polymers, initially 52 structures were downloaded from the database in *Materials Studio*, and then randomly split into training set and test set with the ratio of 70% and 30% (36 and 16), respectively. A defect structure was generated to test the transferability from pristine structures. One elemental crystal (graphite) was added to the training set to increase its complexity, giving a training set with 37 structures and test set with 17 structures.

For the case of zeolites, 5 structures with intermediate size are randomly selected from the database of *Structure Commission of the International Zeolite Association* as the training set. As for the design of test sets, 5 small zeolite structures are manually included to test the transferability from large structures to small while 3 structures larger than that in the training set are also included with similar intention. One defect structure is also manually created to test the transferability from pristine structures.

After collecting structures, for each structure in the training sets, all the symmetrically inequivalent grid-points inside the unit cell with a given spacing (~0.5 Å for polymers and ~0.75 Å for zeolites) are converted into graphs as discussed in the main text. In order to avoid bias towards certain structures, in the pool of graphs from all the structures, the maximum number of graphs from one structure is set be 2000. Then, some graphs are randomly picked from the pool as the training data, on top of which CGCNN is trained. The number of graphs in the final training set is determined by the learning curve shown in FIG. 2(a), for which the convergence criterion is that the difference between the MAE of two trials is less than 0.01 e/$Å^3$.

For the dataset for calculating dipole moments, the four thoroughly studied structures are manually included and other structures are randomly picked, resulting in a set of 13 structures.

For training sets, the grid spacing for polymers is set to ~0.5 Å and for zeolites ~0.75 Å. For test sets, for crystalline polymers and the six zeolites with small unit cells, the charge density is predicted on a grid of ~0.5 Å while for the three large zeolites it is set to ~0.75 Å. For visualization and dipole moments, a refined grid of ~0.25 Å was used.

## 5. Chemical complexity of datasets

For the case of crystalline polymers, in the training set there are 9 elements (C, H, O, N, Cl, F, S, Si, Hg). A simple way to quantitatively evaluate the structure is through a ratio of elements. For example, for molecules with 2 carbon atoms, H:C = 3, 2, 1 indicates a single, double and triple C-C bond, respectively. In addition to H:C ratio [5], C:O ratio is also considered as a descriptor for organic materials [6,7]. In the training set for crystalline polymers, there is a wide range of H:C, from 0.25 to 2.43, and C:O from 1 to 18. Structures in the training set also span a wide range in size, from 8 to 288 atoms in the unit cell. For the test set, the structures are composed of C, H, O, N with the H:C ratio from 0.5 to 2, and C:O from 1 to 7, and the size spans a range from 24 to 504 atoms in a unit cell.

For zeolites, the complexity of structures is lower, with only Si and O atoms and Si:O = 0.5 for almost all structures. The most different structure in the training set is the one with Si:O = 0.48 (SVR), and that in the test set is NPO_defect with Si:O = 0.42, and the difference between the training and test set is mainly in size (120 to 366 atoms and 18 to 576 atoms for the training and test set, respectively).

## 6. List of structures in the training sets

**TABLE S2.** Structures from which training data are obtained for crystalline polymers and zeolites. The last five with 3-digit capital alphabet symbols are zeolites, and others are crystalline polymers.

| name | formula (inside the cell) |
|---|---|
| 1,3-dioxocane | $C_{12}H_{24}O_4$ |
| 1,3-dioxonane | $C_{28}H_{56}O_8$ |
| chloroprene | $C_{16}H_{20}Cl_4$ |
| diketene | $C_{16}H_{16}O_8$ |
| ethO | $C_{56}H_{112}O_{28}$ |
| ethO-HgCl2 | $C_8H_{16}O_4Hg_4Cl_8$ |
| ethO-planar-zigzag | $C_4H_8O_2$ |
| ethoxybenzoate-beta | $C_{72}H_{64}O_{24}$ |
| i-propylene-alpha | $C_{36}H_{72}$ |
| ethylene | $C_4H_8$ |
| gutta-percha-beta | $C_{20}H_{32}$ |
| i-ortho-fs | $C_{114}H_{126}F_{18}$ |
| isopropylethoxide | $C_{20}H_{20}O_4$ |
| i-styrene | $C_{144}H_{144}$ |
| i-vethsilane | $C_{74}H_{180}Si_{18}$ |
| i-vmthether | $C_{54}H_{108}O_{18}$ |
| ketone | $C_{12}H_{16}O_4$ |
| m-xylylene-adipamide | $C_{28}H_{34}O_4N_4$ |
| n-butyraldehyde | $C_{64}H_{128}O_{16}$ |
| nylon-6-10-alpha | $C_{16}H_{28}O_2N_2$ |
| nylon-6-6-alpha | $C_{12}H_{20}O_2N_2$ |
| nylon-7-7-gama | $C_{14}H_{22}O_2N_2$ |
| oxacyclobutane-II | $C_{54}H_{108}O_{18}$ |
| p-phoxide | $C_{24}H_{16}O_4$ |
| p-phsulphide | $C_{24}H_{16}S_4$ |
| p-pht | $C_{28}H_{16}O_8$ |
| p-phtamide | $C_{28}H_{20}O_4N_4$ |
| pppo-alpha | $C_{144}H_{96}O_8$ |
| s-1-butene | $C_{32}H_{48}O_{16}$ |
| s-propylene-2 | $C_{24}H_{32}$ |
| s-styrene | $C_{64}H_{16}$ |
| terylene | $C_{10}H_8O_4$ |
| tetrahf | $C_{16}H_{32}O_4$ |
| trans-dodecenamer | $C_{12}H_{22}$ |
| valcohol | $C_4H_8O_4$ |
| vdenef | $C_4H_4F_4$ |
| graphite | $C_{64}$ |
| AST | $Si_{40}O_{80}$ |
| IWW | $Si_{122}O_{244}$ |
| STT | $Si_{64}O_{128}$ |
| STT_P21c | $Si_{64}O_{128}$ |
| SVR | $Si_{92}O_{192}$ |